# A Novel Acoustic Wearable for Assessment of Tendon Health and Loading Condition

Amirhossein Yazdkhasti[1,2*], Hendrik De Klerk[2], Andreea Renata Lucaciu[2], Rana Moeinzad[2], Hamid Ghaednia[1,2], Joseph H. Schwab[1,2]

[1]Cedars-Sinai Medical Center, Los Angeles, CA, [2]Massachusetts General Hospital, Boston, MA, IN,

[*]amirhossein.yazdkhasti@cshs.org

**Abstract:**

The current methods of assessing tendon health such as clinical examination, imaging techniques, and implanted pressure sensors, are often based on a subjective assessment or are not accurate enough, are extremely expensive, or are limited to relatively large damage such as partial or gross tear of the tendon and cannot accurately assess and monitor smaller damages such as micro tears or strains. This study proposes an acoustic-based wearable capable of estimating tendon load and predicting damage severity in both deep and superficial tendons. Our device consists of an array of acoustic transducers positioned around the targeted body area in the form of a cuff. One of the transducers generates an acoustic wave, which is capable of penetrating deep into the body. As these waves propagate through different tissues, they are influenced by the mechanical and geometrical properties of each tissue. The rest of the transducers are used to measure the propagated waves. The results suggest that the proposed wearable offers a promising alternative to existing superficial tendon monitoring wearable devices by improving the domain of reach. The proposed wearable shows robust performance in estimating the force applied to the tendon. It also can effectively be used to compare the health condition of two tendons and predict the type of damage.

# 1. Introduction:

Tendons are perceived as passive connective tissue between bone and muscles; however, unusual behavior of tendons such as high elasticity and viscoelasticity help us in repetitive motion by storing energy and preventing damage to bone and muscles because of sudden movement. These characteristics alongside nonlinearity and anisotropy make understanding load-related tendon behavior challenging. Moreover, individual variability in tendon properties further complicates our understanding. Factors such as age, gender, and training history can influence the physical performance of tendons[1]. As a result, a non-invasive device capable of measuring tenon loading is highly beneficial across research, clinic, and routine applications.

A significant amount of research has been conducted on estimate or measure tendon loading. Implantable force sensors[2] and Implantable strain sensors [3] are the only devices capable of directly measuring tendon loading; however, they are highly invasive. An indirect method to estimate tendon loading is measuring the elongation of tenon using imaging methods like Magnetic Resonance Imaging (MRI), X-ray imaging[4], computed tomography (CT) scans[5], fluoroscopy[6], and ultrasound imaging [3] and calculating the loading based on the estimated mechanical properties of the tendon. Disadvantages of these methods are limited measurement frequency, small measurement duration, possible radiation, and high operation cost. Another method is calculating the loading using mathematical and biomechanical models based on joint movement and force[7,8] , which requires complex models and assumptions regarding muscular coordination[9].

To address these disadvantages, the better alternative is surface sensing systems. These systems offer non-invasive, real-time, accessible monitoring capabilities with enhanced accuracy. A variety of strategies have been used for this purpose including flexible strain sensing[10,11], shear wave[12–14], and acoustic wave sensing systems[10,11]. A sensing system that relies on tendon deformation is more prone to error since they estimates tendon loading based on deformation measurement and assumed mechanical properties. In contrast, wave-based sensing systems estimate tendon loading by measuring changes in acoustic properties due to applied force. So, it is a more direct measurement method. In these systems, as a wave passes through a tendon, both mechanical and geometrical parameters affect the propagation pattern of the wave. Existing wave-based sensing systems solely focus on mechanical properties and isolate these properties by conducting the measurement locally. As a result, the detection range of these systems is reduced, and they are limited to superficial tendons such as the Achilles tendon.

We believe that both the mechanical and geometrical properties of tendons are crucial. A sensing system capable of detecting both, not only does not suffer from the range limitation of existing sensing systems but also acquires information about the loading, structure, and composition of the tendon. Tendon damage, whether caused by injury or other conditions, changes tendons' functionality, mechanical properties, and structure. Reduction in collagen density and hydration in aging tendons[15], fragmentation and disorganization of tendon collagen fibers due to tendon rupture[16], and changes in size and elasticity of tendon due to tendonitis and tendinosis[17] are examples of these changes. As a result, the proposed sensing system has a potential diagnostic application. Tendon damage is common musculoskeletal injury in both high-level athletes and the general population. Tendon damages account for 30% of all musculoskeletal consultation[18] . Using the proposed system alongside traditional diagnostic techniques like clinical examination functional test, ultrasound imaging, MRI, and XRay to address their limitations such as operator dependency, high cost, and low accessibility.

In this work, we introduce an acoustic-based approach for estimating force applied to tendons and evaluating their mechanical properties and structures. A group of wearable devices are developed to monitor tendons with different musculoskeletal configurations. Figure 1-A illustrates the implementation of the proposed wearable to monitor the Achilles tendon and extensor carpi radialis brevis (ECRB) tendon both of which are prone to damage and have different musculoskeletal configurations. In this paper, we investigate the performance of the proposed sensing system to monitor pig cadaveric tendons. The tendons are tested while sounded by tissue-like material to confirm the long-range capability of the proposed wearable (see Figure 1b). The wearable continuously emits acoustic waves into the artificial limb. As acoustic waves propagate through the artificial limb, the intensity of the wave changes depending on the geometrical and mechanical properties of the tendon and artificial muscle. Then, the propagated wave is recorded by the wearable. We are hypnotized that tension on the tendon has a strong correlation with changes in the acoustic field. Additionally, we explore the potential of the proposed wearable to detect differences in the acoustic behavior of tenons that have been intentionally damaged to mimic various injury patterns and severity.

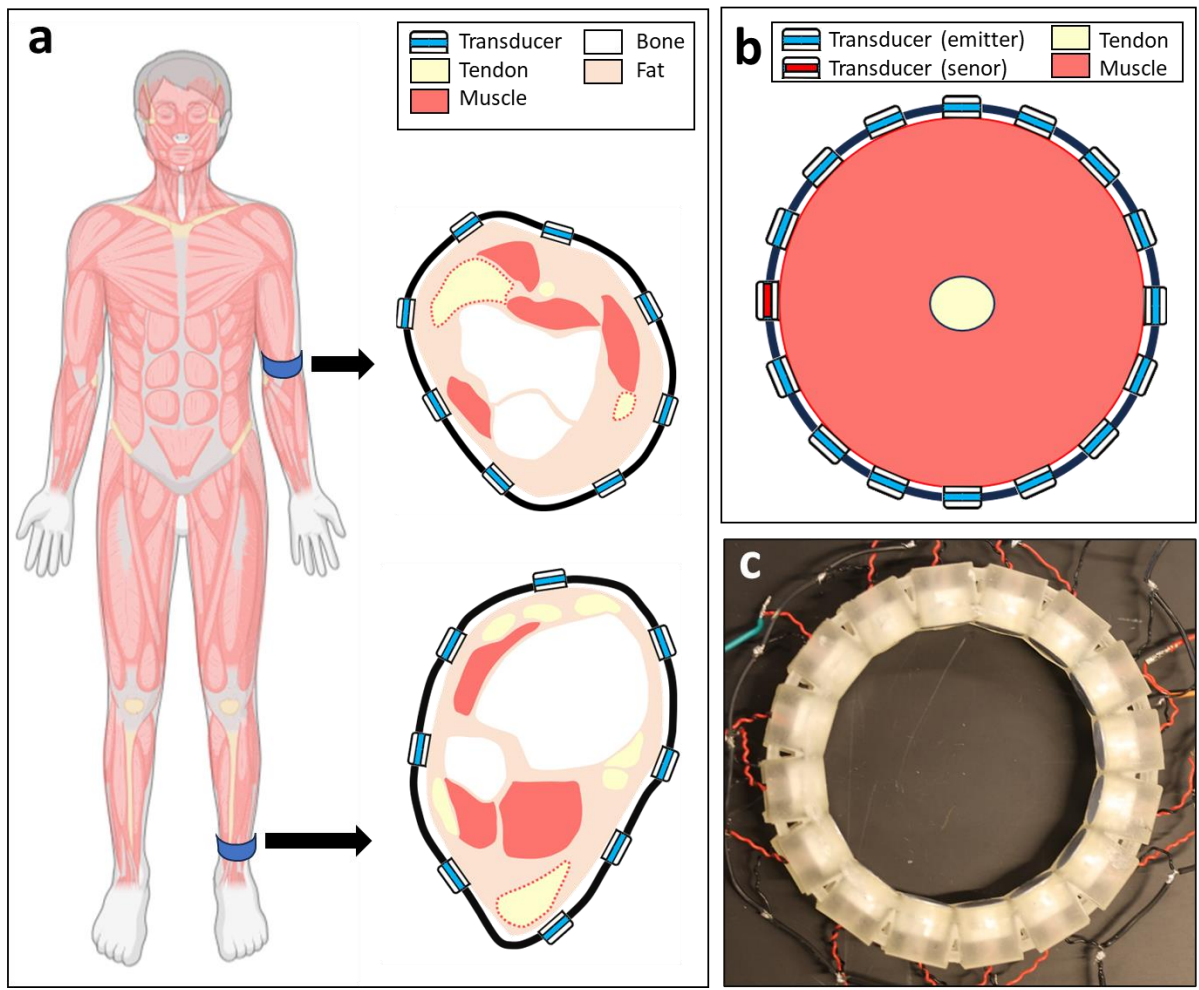

*Figure 1: (A) Schematic of the proposed acoustic wearable used to monitor Achilles and ECRB tendons, (B) Schematic of the experiment setup cross-section (C) photograph of Proposed acoustic wearable*

# 2. results

## 2.1. Design and working principle of the acoustic sensing system.

The Proposed acoustic wearable Is a 16-element array arranged in the form of a cuff which can be worn on the limb around the target tendon. Each element is an ultrasound transducer capable of generating and measuring acoustic waves in the frequency range of 47kHz to 57 kHz. These transducers have the derated spatial-peak temporal-average intensity of ($I_{SPTA.3}$) 25.8 mW/cm$^2$ , so they comply with the Food and Drug Administration (FDA) guidelines on acoustic diagnostic systems [18]. They also have an aluminum casing, so they are easy to clean and disinfect, making them suitable for wearable use.  A 3D-printed structure made of elastic material is used to fix the relative position and orientation of transducers. This structure is optimized to be conformable, durable, and stable. A sufficient contact force between transducers and skin guarantees a strong acoustic field and stable signal measurement. At the same time, a large contact force leads to discomfort and short-lived wearables. An elastic 3d printed structure is designed to maintain this balance.

In each burst, one of the transducers generates a two-cycle sinusoidal signal with a frequency of 52 kHz.  As this acoustic pulse passes through different tissues, its propagation pattern changes depending on the mechanical and geometrical properties of these tissues.  These changes are recorded by other transducers, allowing us to extract information about the alteration of these properties from variations in the recorded signal. In a complex movement, source of changes in acoustic field can be changes in bones[], muscles, and tendons. To extract information about the tendon specifically, we need to isolate the movement of the tendon in such a way that it is the only changing tissue. In this scenario, the tendon experiences elongation under axial load. This elongation resulting in shrinkage in the cross-section area according to Poisson's ratio[19].At the same time, due to nonlinearity of tendon, The stiffness of tendon changes because of the loading. All of three these factors contribute to changes in acoustic behavior of teens while  the magnitude of changes due to each of factors is unclear , thus further analysis is necessary,

Generally, collagen fibers are organized alongside the tendon. So, despite the variation, the overall mechanical property, structure and composition of the tendon is assumed to be constant along the tendon. Moreover, the uniformity of cross section along the tendon is a reasonable approximation, based on the dimension measurement of tested tendons. Consequently, the elongation of tenon locally displace the tendon; however, it can not significantly change the acoustic field. Separating contribution of transvers deformation and stiffness change on tendons acoustic behavior is practically impossible since you can not have one without the other one. Finite element method (FEM) models have been used to investigate the effect of each of these phenomena independently.

Although we envision using the wearable for different tendons, the dissected tendon has a cross-section close to the Achilles tendon[20]. Three rounds of simulation are conducted based on elongation and stiffness changes of the Achilles tendon during repeated hopping exercises reported in the literature[21]. It is shown that during a ramp contraction with the muscle force of 60 N, the tendon experienced 5 mm of elongation and a relative change in stiffness of 20%[21].In the first round of simulation, the shrinkage of tenon cross-section with the assumption of uniform material and simplified geometry is calculated through 3d simulation. The calculated cross-sectional is used to simulate the acoustic field when only geometrical properties affect the acoustic behavior. In the second round of simulation, the acoustic field is simulated under similar geometrical conditions while relative stiffness is gradually increased to 20%. To investigate the contribution of geometry and stiffness, in the last round of simulation, both geometry and stiffness are changed simultaneously.

Figure 2-a shows changes in the acoustic field differ when only geometry changes, only stiffness changes, or both parameters change simultaneously under 52 kHz acoustic actuation. This indicates that both geometry and stiffness are contributing factors in acoustic propagation within the tendon. We also investigate the effects of these changes on surface acoustic measurements to evaluate whether the proposed wearable can detect the effects of both factors.

Figure 2-b illustrates the relationship of surface acoustic measurement with tendon deformation and stiffness at eight locations when one or both parameters are altered. The figure suggests that both geometry and stiffness influence the superficial acoustic measurement; however, the magnitude of these changes highly depends on the measurement location. At locations of 90° and 165 ° relative to the transmission direction, geometrical changes have more contribution while at 15 °relative to the transmission direction, stiffness contributes more to acoustic measurement. Therefore, superficial acoustic measurement at different locations is essential to capture the effects of both geometrical and mechanical factors.

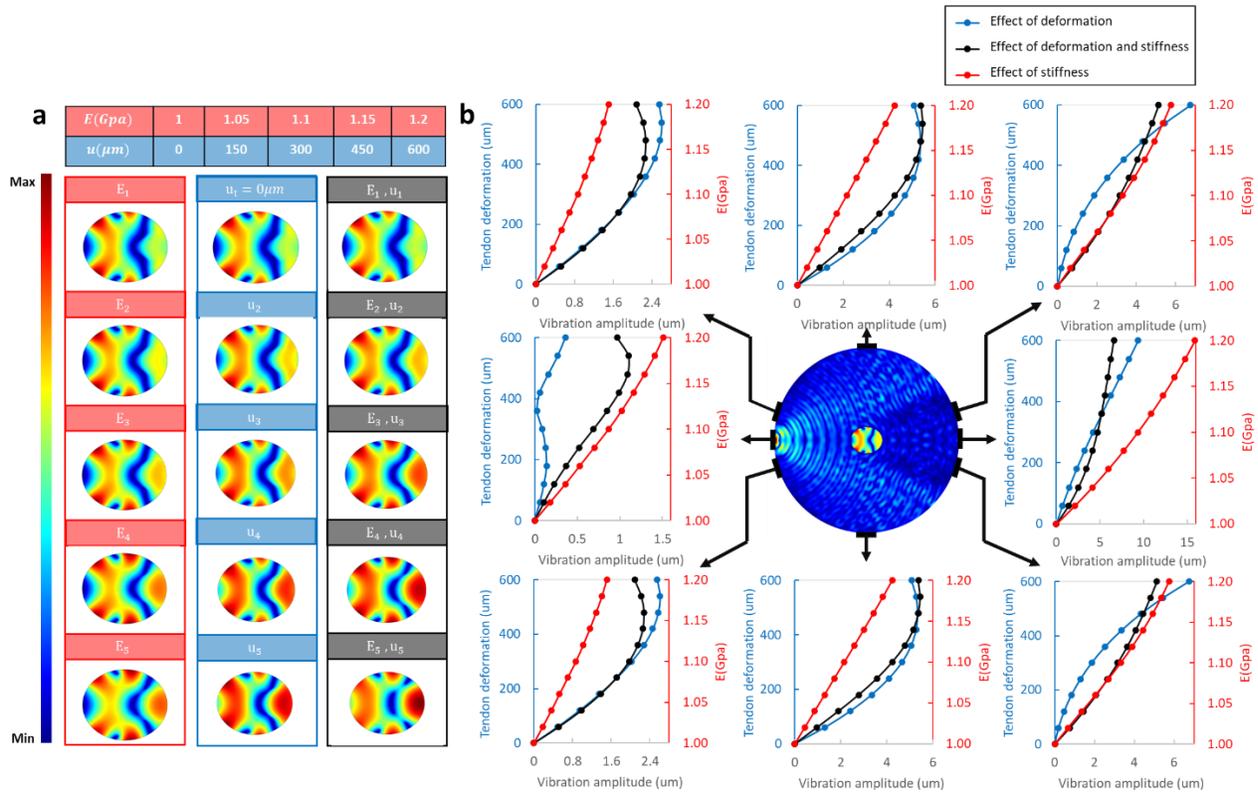

*Figure 2: (a) Simulated displacement field of the tendon in response to variations in stiffness, deformation, and their simultaneous occurrence. (b) Simulated changes in surface vibration caused by variations in Young's modulus, tendon deformation, and the combined effects of both.*

## 2.2. Mechanical testing.

To investigate the performance of the proposed sensing system knowing the true value of load or damage is essential. Since there isn't any non-invasive method to accurately measure tenon load on living subjects, the experiment we conducted on pigs' cadavers. The porcine flexor digitorum of

18 pigs' cadaveric fresh-frozen legs is used in this study. The tendons are dissected and cut into 8cm specimens. These specimens have an average weight of 8.5 g and an average cross-section area of 92.36 mm$^2$. The harvested tendons are temporarily stored in a physiological saline solution with a temperature of 36 to maintain hydration.

In addition to healthy tendons, the experiment was conducted on damaged tendons. Three categories of damage were artificially induced in the tendon including partial longitudinal rupture, partial transverse rupture, and micro tears. Partial tears are induced by longitudinal and transverse cuts with a depth of 20 % of tendon thickness. Microtears are formed by a one-time application of forces beyond the tendon's functional limits, leading to small damage to the tenon's internal structure.

To prevent dehydration and provide an environment similar to the body, harvested tendons are placed in the center of a cylindrical block of muscle-like gelatin with a diameter of 7 cm, see Figures 3-a and 3-c. This gelatin has acoustic properties similar to the muscles. The proposed wearable is wrapped around the gelatin block and continuously generates and records acoustic waves. Two ends of the tendon are fixed by textured grips to the test setup. One end is connected to a force sensor and the other end is connected to a high-accuracy linear stage. A signal generator (DG822 Arbitrary Waveform Generators, RIGOL Technologies) is used to produce the signals, a data acquisition system(PicoScope 4824A, Pico Technology ) is used to record the data from the cuff, and the force sensor ( FC 2k ZEUS, Scientific Industries Inc.) was used to measure the applied load in real-time.

The trend undergoes cycles of stretching and unloading within the elastic domain. To find the maximum load in which tendons show elastic behavior, a pilot study has been conducted. Multiple experiences on different tendons show that applying force beyond 60 N causes damage which is defined as an irreversible change in the elastic behavior of the tendons[22]. Details of our pilot study are shown in the supplementary figure??? . Another important parameter of loading that needs to be optimized is the number of cycles. The larger number of cycles gives us insight into consistency, repeatability, and measurement noise. On the other hand, a large number of cycles leads to a decline in the mechanical properties of non-living tendons. Some studies have recommended exposing tendons to less than six cycles of loading and unloading[23] . As a result, in this study, tendons undergo 5 cycles of stretching and unloading with a maximum load of 60 N. Each cycle takes 10 min, and the force and acoustic signals are recorded after each .2 mm deformation of the tendon.

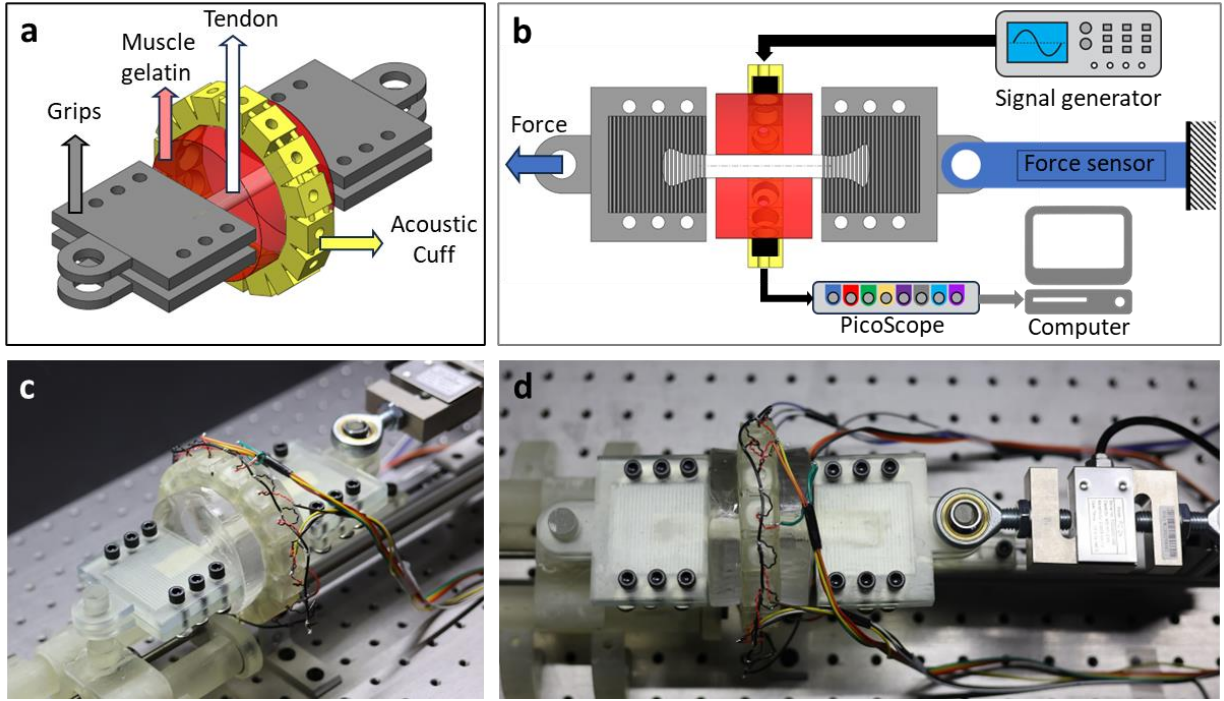

*Figure 3: (a) Schematic of the test sample consists of a dissected tendon surrounded by gelatines. Both sides of the tendon are fixed by gripes, and a 16-element acoustic wearable raped around the sample. (b) Schematic of the testing setup consists of the test sample, data acquisition system, signal generator, and force sensor. (c, d) photograph of the test sample and testing setup.*

### 2.3. Results processing.

At the interface of muscle and tendon, different acoustic phenomena happen, depending on the relative acoustic properties of the muscle and tendon. Part of the emitted wave is reflected at the interface of the tendon, while another portion is refracted. At the same time, part of the wave is transmitted through the tendon. The interference of these three phenomena forms the acoustic field. To address hardware limitations and reduce the computational cost, we process signals from 8 out of the 16 elements. The positions of eight selected elements are chosen to allow us to investigate the role of each of the mentioned acoustic phenomena in changes in the acoustic signal caused by variations in the tendon's mechanical properties.

Noise is an inevitable part of experimental measurement, particularly in acoustic systems. existence of environmental and electronic noise potentially caused undesirable effects on the accuracy and reliability of the proposed system. A band-pass filter is implemented to manage these noise sources. To reduce the data complexity and simplify the analysis, the Hilbert transformation is used to calculate the envelope of the recorded signals. The filtered and enveloped signals measured by different transducers are shown in Figure 4-$a_{1-8}$ with black and red solid lines. These Figures show that the profile of the measured signal is highly affected by the location of the measurement. Moreover, the maximum measured acoustic intensity in Figures 4 $a_1$, $a_2, a_6$, and $a_5$, is higher than in Figures 4 $a_3$, $a_5$, and $a_8$. This suggests that the portion of the acoustic energy reflected from the tendon is larger than the portion of acoustic energy transmitted through it, which aligns with the simulated acoustic field shown in Figure 2-b.

Visualizing and interpreting changes in acoustic signals for different tendon deformation can be challenging, so all of the signals recorded by a single transducer during the loading or unloading process are combined into a spectral map. In these spectral maps, each horizontal line represents the normalized enveloped acoustic signal at the specific deformation point, and each horizontal line indicates the variation of acoustic intensity in a time instance at different tendon deformation. As illustrated in Figure 4-b, an increase in deformation leads to variations in the acoustic signal at different time points, reaching its maximum at 0.37 ms.

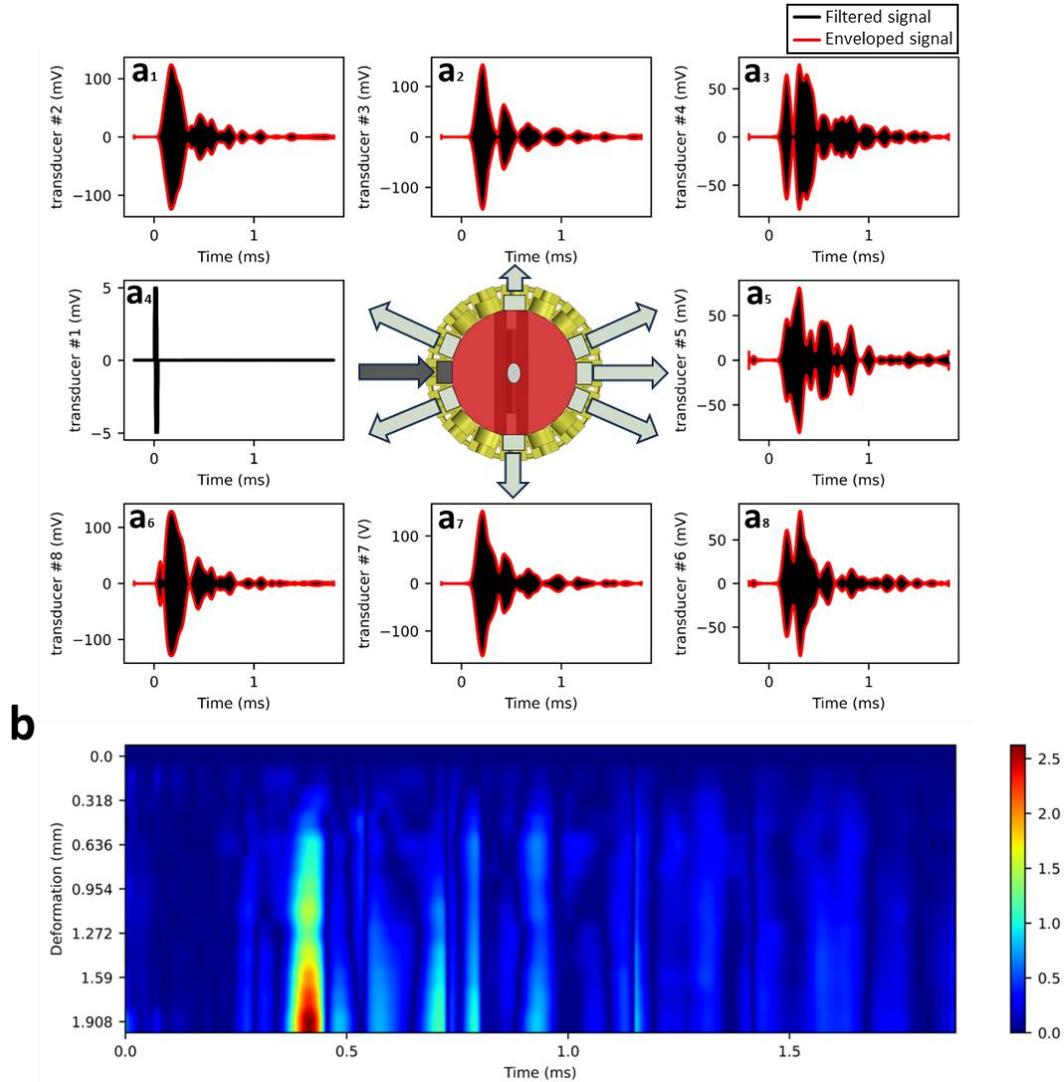

*Figure 4: ($a_{1-3}$ and $a_{5-8}$) Filtered and enveloped acoustic measurement by transducers located at different positions. ($a_4$) Actuation signal for the transmitter responsible for generating acoustic field (b) Spectral maps in the tendon deformation versus time of measured signal by the transducer positioned in front of the actuated transducer.*

### 2.4. Stress-Strain behavior.

The mechanical properties of tendons under physiological stress are strongly influenced by the quality, quantity, and alignment of collagen fibers, as well as the integrity of the extracellular matrix (ECM)[24] . Although variations in collagen synthesis, fiber orientation, and ECM composition are inevitable among specimens dissected from different animals, the extent of their impact on tendon mechanical properties remains unclear. Therefore, investigating the variance in mechanical properties across different specimens is crucial. So, tensile tests are conducted on the 18 healthy tendons. Variation in fiber crimp patterns within the tendons is also a source of inconsistency[25]. To minimize the effect of this variability, a 5N pre-stress is applied to tendons prior to the tensile test. Another important aspect is the viscoelastic nature of tendons[26]. In the proposed system, recording data is time-consuming. So, the time window of measurement is limited to 1 minute for each deformation, and the force over this period is averaged to account for the time dependency of the viscoelastic tendon.

As a tendon is damaged, the quality and alignment of collagen fibers is compromised which leads to changes in mechanical properties. The amount of these changes depends on the magnitude and distribution of destruction in the tendon structure. The longitudinal partial rupture, transverse partial rupture, and microtear cause discontinuity in collagen fibers and compromise their alignment, each to a different extent. In partial transverse rupture, approximately 15% of fibers are discontinued, so stress is more concentrated on the rest of the fibers. In partial longitudinal rupture, although some fibers are discontinued and their orientation changed, the number of affected fibers is significantly smaller than in transverse rupture. In microtear, the ruptured fiber is distributed in a large area. These differences in magnitude and distribution may have different effects on mechanical properties that can be measured by tensile test. So, a tensile test is conducted on 6 damaged tendons for each damage scenario.

The stress-strain curves of loading and unloading are shown in Figures 5-$a_{1-4}$ and 5-$b_{1-4}$. To analyze the average stress-strain behavior of tendons across different scenarios, the stress-strain curves from repeated measurements on various tendons were averaged and plotted in Figure 5-$c_{1-4}$. To compare Stress-strain curves of different tendons, we can focus on two key features: stress amplitude and overall profile of loading and unloading curves. The comparison of stress amplitudes cannot be used to classify damage since it depends on both inherent variation in tendon and the extent of damage. On the other hand, tendons with different damage conditions show difference in overall profile of loading and unloading curve which can be used to classification of damages

The area under the stress-strain curve is the energy of mechanical deformation per unit of volume. This area can be used to compare stress-strain behavior of tendons in different scenarios. The energy density required for loading and the energy density released by unloading for different scenarios are shown in Figure 5-$d_{1-2}$. Another important aspect to consider is hysteresis. Hysteresis represents the energy density dissipated in the loading and unloading process. Hysteresis can be derived from the area between the loading and unloading curve. Due to variation and heterogeneity in tendon behavior, we compare the rate of dissipated energy to the energy required for loading across different damage scenarios which is shown in Figure 5-$d_3$. As this figure demonstrates, the relative dissipated energy of damaged tendons is on average 9.3 % lower than healthy tendons. Among the damaged tendons, the transvers rupture shows the lowest dissipated rate which can be justified by discontinuation of group of fibers and reduction in cross section due to the damage

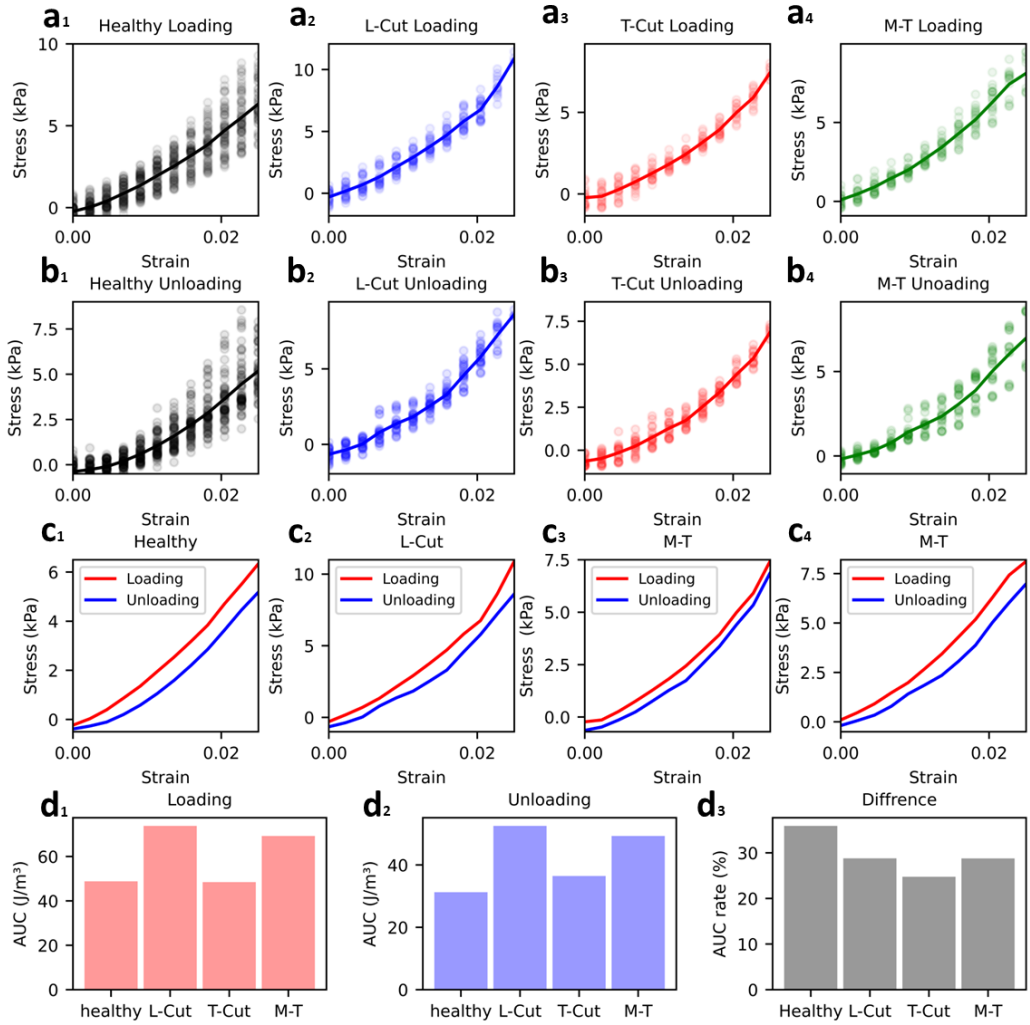

*Figure 5: ($a_{1-4}$) Loading Stress-Strain relation of healthy, Longitudinal rupture, transverse rupture, and micro tear: Mean Stress-strain curve(solid line) and Stress-Strain of all the cycles of all the tendons (dotted line) ($b_{1-4}$) Unloading Stress-Strain relation of healthy, Longitudinal rupture, transverse rupture, and micro tear: Mean Stress-strain curve(solid line) and Stress-Strain of all there the cycles of all the tendons (dotted line). ($c_{1-4}$) average stress-strain of a cycle. ($d_{1,2}$) Area under the averaged Stress-Strain loading and unloading curves in different damage scenarios. ($d_3$) relative difference in AUC of loading and unloading in different damage scenarios.*

### 2.4. Stress-Acoustic behavior.

Additional to stress measurement, the acoustic behavior of tendons are recorded during the tensile test. As it is mentioned, 18 healthy and 18 damaged tendons undergo 5 cycles of loading and unloading. In total, 2520 pairs of spectral maps and force measurements form our database. Comparing the behavior of different tendons and drawing meaningful conclusions for such a large dataset is challenging. To simplify the analysis and interpretation, we reduce the dimensionality of the data by averaging the envelope of the signal at each deformation. As a result, each spectral map is summarized into one 2-dimensional plot that can be compared with force measurement. Although we lose the time domain and frequency domain information, the overall average intensity trends are highly correlated with the loading of the tendon.

Acoustic-stress curve of loading and unloading of different tendons measured by transducer located in front of actuated transducer are shown in Figure 6-$a_{1-4}$ and 6-$b_{1-4}$. Generally, variance in healthy tendon is larger than the damaged tendons. Source of this variation can be the larger number of tendons in healthy class and dissimilarity in geometrical and mechanical properties of tendons. As mentioned, the acoustic intensity measured by each transducer not only depends on the geometrical and mechanical properties of the tendon, it also depends on the location of the transducer. Despite these differences, the trend of increase in average acoustic signals is observed in all of the transducer's measurements. Figure 6-$c_{1-4}$ illustrates the average of acoustic-stress curve of 7 transducers if different scenarios for loading and unloading. The margin of variance between different transducers is shown by shaded area for each scenario.

At first glance, the stress-strain curve and stress-acoustic curve are similar to some extent. Both have positive slop with comparable values, and this slop increases in higher stress levels which leads to positive concave. So, the correlation between force and average acoustic intensity follows the correlation between force and deformation. It does not necessarily mean that acoustic intensity can be used as a direct measurement of force; however, the application of force modifies certain mechanical and geometrical parameters that influence tendon interaction with acoustic waves. This enables the proposed wearable to estimate tendons loading.

A more detailed comparison between stress-strain and stress-acoustic curves shows a more distinct correlation of stress- acoustic overall pattern of loading and unloading with damage induced to the tendon. The Longitudinal rupture and healthy tendons have the most similarity to each other which is aligns with expectations, as the longitudinal rupture has minimal effect on both geometrical and mechanical properties of the tendon. In case of transvers rupture, the difference between loading and unloading is generally smaller than other scenarios, especially in higher stress levels. This observation might be reasonable, as no force is applied to the ruptured area, so changes in mechanical properties of these are less sensitive to force. This can fade the difference of acoustic behavior between loading and loading. In the micro tear scenario, the curvature of the stress-acoustic curve is lower than other scenarios. This indicates that the rate of changes in acoustic intensity is more influenced by stress level. Since this trend is not observed in stress-strain curves, it suggests that formation of micro tears has less noticeable impact on the integrity of tenon compared to its effect on acoustic behavior. All in all, differences in stress-acoustic curves among different damage scenarios are more visible compared to stress-strain curves.

Another way to compare stress-acoustic behavior of tenons in different scenarios is comparing area under the load and unlading curves, see Figure 6-$d_{1-2}$. The relative difference in AUC of loading and unloading, shown in Figure 6-$d_3$, confirms our observation in stress acoustic curves. Healthy and longitudinal rupture show similar behavior while transvers rupture and microtears show lower difference with different extends. The similarity of health and longitudinal rupture resulted in minor changes in mechanical and geometrical properties of tendon because of longitudinal rupture. Transvers rupture and micro ruptures followed by the discontinuation of a group of fibers, cause the acoustic behavior of tendons with these conditions diverge from that of health tendon. This divergence along with the overall profile of stress-acoustic curve are effective tolls for classifying damage type and severity of a tendon.

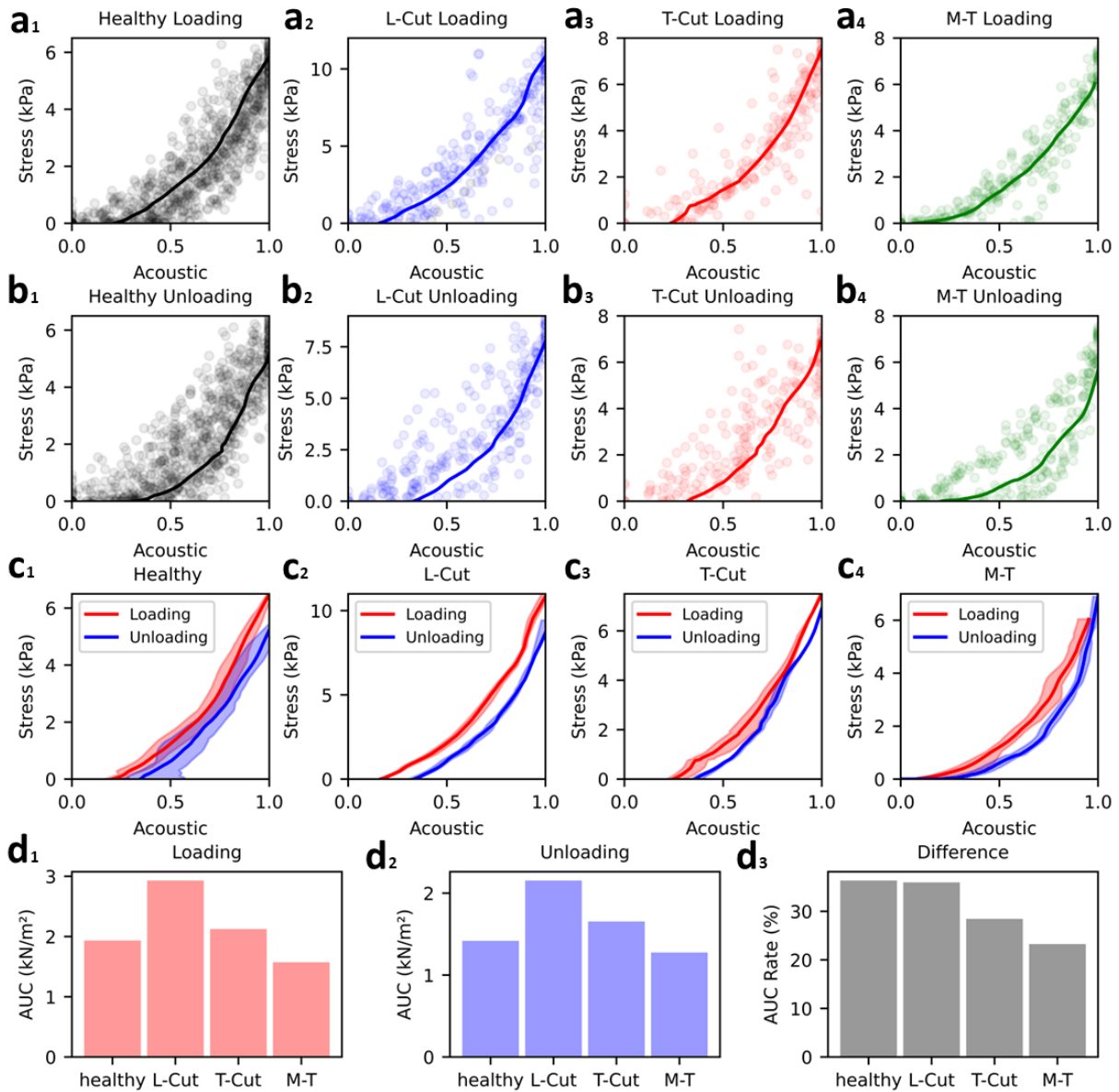

*Figure 6: ($a_{1-4}$) Loading Stress-Acoustic relation of healthy, Longitudinal rupture, transverse rupture, and micro tear: Mean Stress-Acoustic curve(solid line) and Stress-Acoustic of all the cycles of all the tendons (dotted line) ($b_{1-4}$) Unloading Stress-Acoustic relation of healthy, Longitudinal rupture, transverse rupture, and micro tear: Mean Stress-strain curve(solid line) and Stress-Acoustic of all there the cycles of all the tendons (dotted line). ($c_{1-4}$) Stress-Acoustic statistical parametric mapping of healthy tendons, tendons with longitudinal rupture, transverse rupture, and micro tears for different transducers. ($d_{1,2}$) Area under the averaged Stress-Acoustic loading and unloading curves in different damage scenarios. ($d_3$) Relative difference in AUC of loading and unloading in different damage scenarios.*

## 4. Discussion and future work

There are two salient findings from this study that enable the use of acoustic waves to monitor tendons, during human movement. First, modeling and Isolated tissue studies revealed that the acoustic behavior of tendon is highly influenced by loading of the tendon. Second, different damaged artificially induced to the tendon changes its acoustic behavior statistically. Despite tendon heterogeneity, we found a strong direct relationship between force applied to of tenon and changes in acoustic behavior of pig cadaveric tenon. Further human studies are essential to investigate the performance of proposed wearable in prediction of static and dynamic loading of tendon compared to gate analysis and imaging alternatives and devices.

The proposed wearable can measure trends in force applied to the tendon. This capability is especially valuable for monitoring time-dependent force and comparing the loading pattern of two tendons present in two limbs. While the wearable excels at detection variation in loading, it currently cannot predict the exact force value applied to the tendon. To enable quantification of these forces, a calibration protocol must be developed.

In this study, we revealed that damage induced in cadaveric pig tendon changes their acoustic behavior under small loadings. Considering that, changes in acoustic behavior of living tissue should be more drastic due to inflation following the damage, routine tenon monitoring could be crucial for diagnosis of tendon injury. Furthermore, our findings indicates that tendons with different induce damage condition, exhibit statistical differences in their acoustic behavior. This shows a great potential of proposed system to predict the severity and nature of tendons damage. we are envisioning number of applications, from injure prevention in aging population and athletes, to aiding rehabilitation and managing condition such as Achilles tendonitis and rotor cuff tendinitis, each of which need their own clinical trial.

We have developed and demonstrated an integrated acoustic sensing wearables that offer noninvasive continuous measurement of changes in mechanical and geometrical properties of a targeted tendon. These wearables can be used to capture information about the dynamic and static loading of the tendon and its health condition. The proposed wearable is used an array of transducer to simultaneously generate and measure acoustic waves at the frequency of 52kHz. To simplify the experiment, only one transducer is used to generate acoustic signal. Using a broader frequency domain, and different excitation pattern, enable us to improve our understand about the physiology specially working with a more complex scenario.

With the aim of simplifying data analysis, the average of the enveloped acoustic signal is used to describe acoustic behavior. Consequently, a significant portion of information embedded in the signal is neglected. As a result, exploring alternative parameters to describe acoustic field is an essential next step. An example of these parameters is viscoelasticity of tendon that can be extracted from the real-time acoustic behavior. Changes in tendon viscosity are an indication of diabetes [23] .So, the proposed wearable and be used a diagnostic tool for diseases similar to diabetes that do not damages tendon but affect it properties.

# 5. Method

**5.1. Numerical Modeling.**
We used the "Pressure Acoustics, Frequency Domain" module in COMSOL Multiphysics software version 5.4 (COMSOL Multiphysics®, Burlington, MA, USA) to develop a simplified numerical model of propagation of acoustic wave in our experimental phantom consists of cadaveric tendon under loading surrounded by artificial muscle. Although the tendon has a 3-D geometry, we model a cross section of the 3-D domain in a 2-D environment. The tendon is model as an ovel with longest diameter of 12 mm and shortest diameter of 9mm covered by artificial muscle with the diameter of 7 cm. A 2cm transducer mounted on the surface of artificial generates 52 kHz sinusoidal wave. The area that is not covered by the transducer is assumed to have low-reflective boundary and all the interior boundary assumed to be fixed. The entire model is meshed with an extremely fine triangular elements which has a maximum element size of 200 μm and minimum element size of 14 μm, as a result, the selected mesh consists of 296208 elements.

Accurate assignment numerical value to mechanical properties of muscles and tendons is inherently challenging since there is compositional and structural variability both within a single person and among different individuals. At the same time, there is an ongoing fundamental debate on some mechanical properties such as Poisson's ratio. In most studies it is assumed to be positive value between 3.5-5 [28], while in few studies, it claimed to be a negative number [29]. Despite these variability, mechanical properties of the materials used in the model are given in Table 2 based on literature review.

| Material | Density (kg/m3) | Young's modulus (pa) | Poisson's ratio |
|---|---|---|---|
| Muscle | 1090 | 0.076e9 | 0.4 |
| Tendon | 1109 | 1.1e9 | 0.42 |

Table 2. Mechanical properties of different tissues used in numerical simulations.

### 5.2. Specimens' preparation.
The porcine flexor digitorum of 18 pigs' cadaveric fresh-frozen legs is used in this study. The tendon is dissected distally from the bifurcation toward the most proximal part of porcine leg (see Figure 7-a). After excising the skin and connective tissues over the flexor digitorum muscle and tendon, we start the extraction by increasing the point where the tenon is splitting into four tendons. Then the tendon is separated from the surrounding tissue using blunt dissection tools. We continue separating the tendon distally until 9 cm of tendon is obtained.

The cross-sectional area of the dissected tendon is measured by capturing images form two directions: one along the tendon's largest diameter and the other perpendicular to it. Assuming the tendons has an ovel cross-section, the images are analyzed with respect to an external length unit, to calculate the area of different cross section of the tendon. An average of these cross-section areas is assigned to each tendon's area. The detailed weight and cross-section measurement for all eighteen tendons used in this study is shown in Table 1 supplementary material.

In addition to healthy tendons, the experiment was conducted on damaged tendons. Three categories of damage were artificially induced in the tendon including partial longitudinal rupture, partial transverse rupture, and micro tears. Transvers rupture is induced by making a transvers incision at the midpoint of the tendon. The largest diameter of tendon midpoint is measured using a caliper. The caliper is then used to mark the incision depth, which is 20 % of measured diameter. Finally, the marked area is carefully incised using a scalpel blade (Figure 7-d). For the longitudinal rupture, a similar process was followed, but incision was made along the tendon, targeting the tendon's center instead of the side (Figure 7-e).

Microtears are formed by one time application of forces beyond the tendon's functional limits, leading to small damages to the tenon's internal structure. The applied force is c

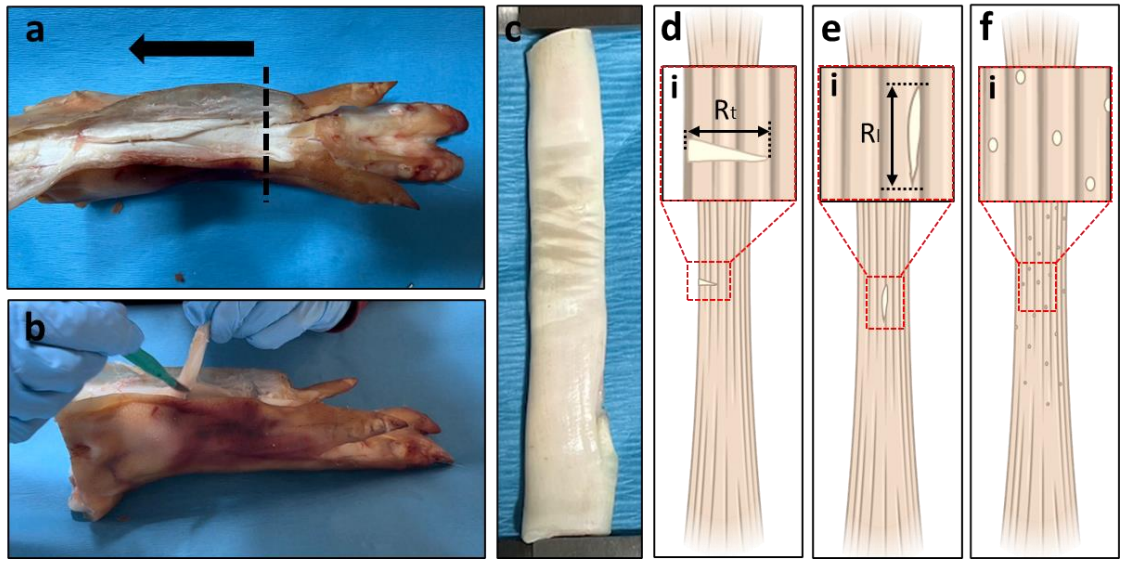

*Figure 7: (a) Palmar view of a porcine leg showing a longitudinal incision on the skin overlying the flexor digitorum profundo's muscle and tendon. (b) Dissection process: the tendon is carefully isolated from surrounding soft tissues using blunt dissection tools. (c) Example dissection tendon. (d-f) Schematics of tendons with transvers rupture, longitudinal rupture, and micro tears*

### 5.3. Transducer characterization.

experiment is conducted on directivity and frequency response of transducers in a customized deionized water tank to isolate environmental noise and undesired reflections. The pressure generated by the transducer is measured by a broadband reference hydrophone (4mm needle hydrophones, Precision Acoustics Ltd) along the access of the transducer at the distance of 1 cm from the transducer. The measured data was recorded using a data acquisition system (PicoScope 4824A, Pico Technology ). The frequency response is obtained by frequency sweeping actuation signal form20 kHz to 70 kHz with a resolution of 100 Hz which is shown in Supplementary Figure S1-a. The peak performance of transducers occurs at 52.2kHz with the intensity of 25.1 mW/cm$^2$. The directionality of the transducers is obtained by measuring the acoustic pressure generated by transducer along line perpendicular to transverse axis at distances of 1.5 and 7 cm while transducer is transmitting 52 KZ wave. Supplementary Figure S1-b and c shows that the acoustic main beam of the transducer is wide enough to cover tendon and its surrounding area.

**Authors and contributions:**

Amirhossein Yazdkhasti: development of the idea, experiment, processing, and writing
Hamid Ghaednia: development of idea, experiment, and writing
Joseph Schwab: development of the idea and writing
Hendrik De Klerk: Sample preparation and experiment
Andreea Renata Lucaciu: Experiment

**Supplementary methods:**

Table s-1 : geometry and mass measurements of all of tendons

| Exp # | Condition | X (mm) | Y(mm) | Z(mm) | Area($mm^2$) | $m_{intial}(g)$ | $m_{end}(g)$ |
|---|---|---|---|---|---|---|---|
| 1 | LC | 9.9 | 12.8 | 80.8 | 99.4752 | 7.778 | 6.92242 |
| 1 | TC | 8.1 | 11.6 | 81.5 | 73.7586 | 6.938 | 6.17482 |
| 1 | MT | 9.6 | 10.9 | 82.1 | 82.1424 | 8.861 | 7.0888 |
| 2 | LC | 8.7 | 11.2 | 83.5 | 83.3199 | 7.212 | 5.7696 |
| 2 | TC | 7.5 | 10.25 | 80.35 | 76.43938 | 6.678 | 5.3424 |
| 2 | MT | 6 | 12.4 | 80.25 | 80.7922 | 5.48 | 5.3 |
| 3 | LC | 7 | 12.2 | 86 | 67.039 | 6.167 | 5.454 |
| 3 | TC | 8.8 | 15 | 80 | 103.62 | 7.433 | 6.988 |
| 3 | MT | 7.44 | 12.77 | 80 | 74.58191 | 6.085 | 5.858 |
| 4 | MT | 7.2 | 13.9 | 82.5 | 78.5628 | 8.126 | 6.98 |
| 4 | TC | 7.3 | 13.5 | 89.8 | 77.36175 | 7.727 | 6.637 |
| 4 | LC | 7.8 | 11.5 | 80.2 | 70.4145 | 9.642 | 7.178 |
| 5 | LC | 6.35 | 13.2 | 82.4 | 65.7987 | 5.271 | 4.141 |
| 5 | TC | 10.6 | 13 | 81.8 | 108.173 | 11.885 | 8.373 |
| 5 | MT | 15.5 | 11.9 | 82.6 | 144.7933 | 12.934 | 6.896 |
| 6 | MT | 11.15 | 17.35 | 91.25 | 151.8602 | 9.393 | 7.411 |
| 6 | TC | 11.3 | 16.35 | 92 | 145.0327 | 10.316 | 7.232 |
| 6 | LC | 10.98 | 10.1 | 78.9 | 87.05493 | 12.061 | 9.425 |

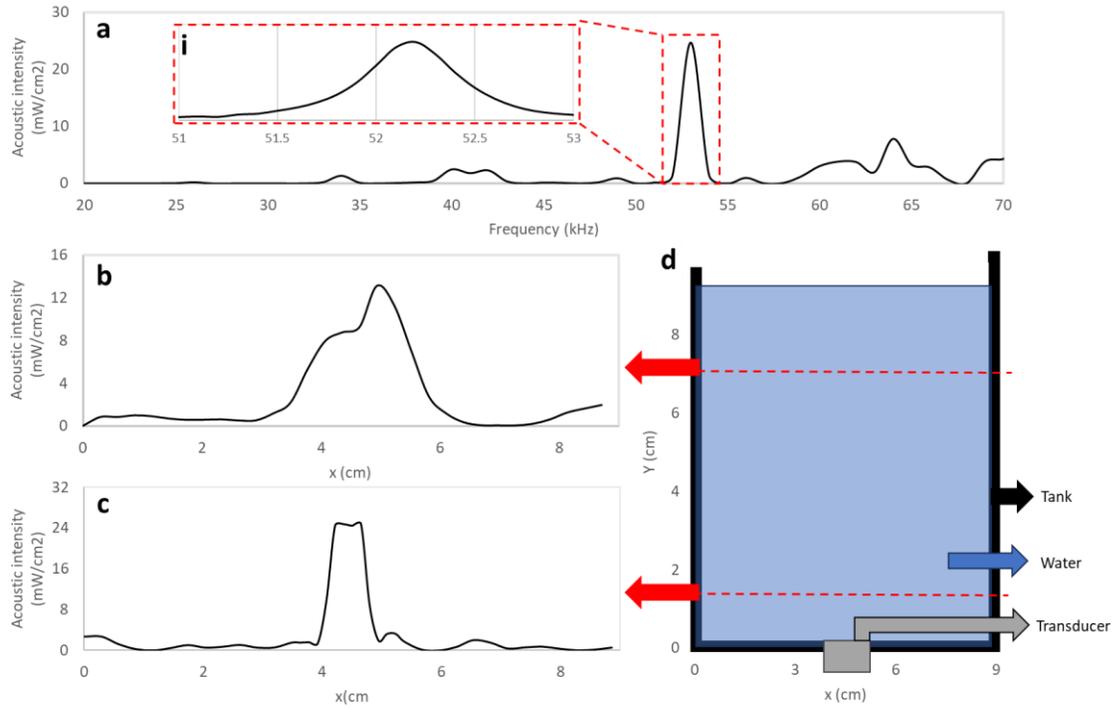

Figure s-1: Transducer characterization: (a) frequency response of the transducers, (b, c ,d) Directivity pattern of the transducer

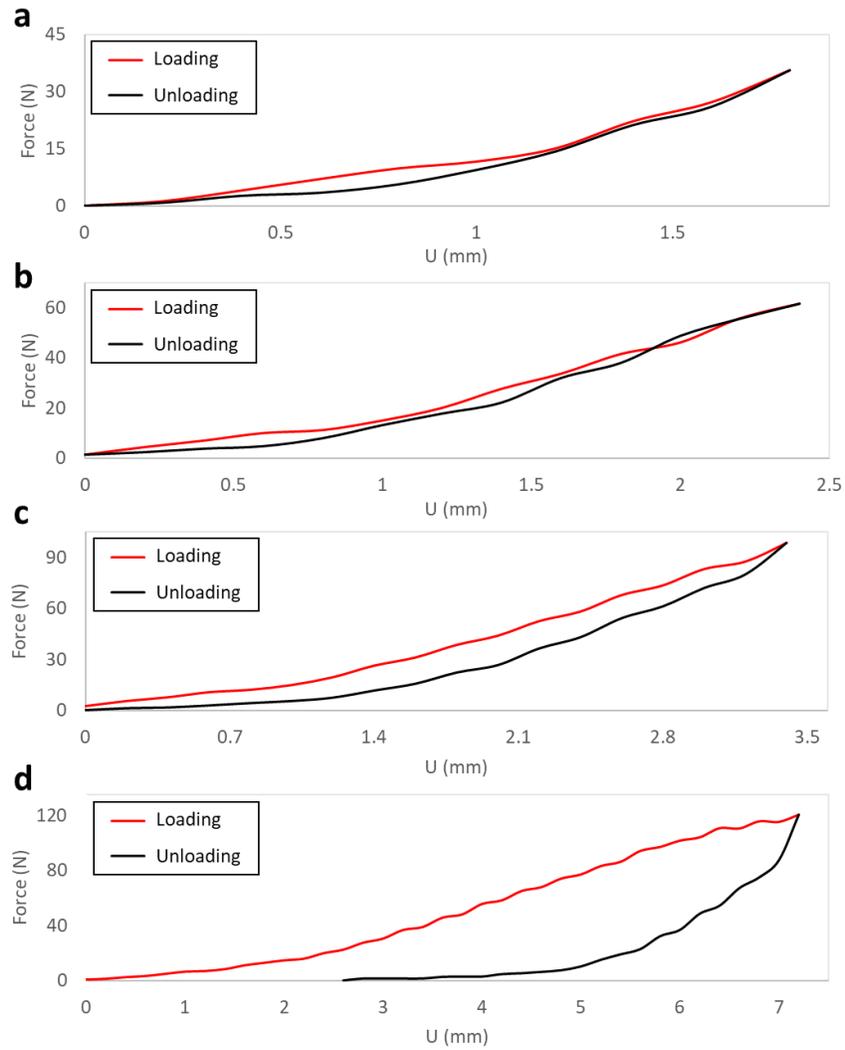

Figure s-2: Finding elastic threshold of tendon: force-deformation behavior of tendons in a loading and unloading cycle with the maximum force of (a)30, (b)60, (c)90, and(d)120